\documentclass[apj]{emulateapj}
\usepackage{color}

\def\gtorder{\mathrel{\raise.3ex\hbox{$>$}\mkern-14mu
             \lower0.6ex\hbox{$\sim$}}}
\def\ltorder{\mathrel{\raise.3ex\hbox{$<$}\mkern-14mu
             \lower0.6ex\hbox{$\sim$}}}

\usepackage{graphicx}	% Including figure files
\usepackage{amssymb}	% Extra maths symbols
\usepackage{colordvi}
\usepackage{amsmath}
\usepackage{hyperref}

\usepackage{academicons}

\slugcomment{Accepted to AJ}

\shorttitle{Asteroid collisions}

\shortauthors{Ofek et al.}

\begin{document}

\title{Asteroid collisions: expected visibility and rate}
\author{
Eran~O.~Ofek\altaffilmark{1},
David~Polishook\altaffilmark{2},
Doron~Kushnir\altaffilmark{1},
Guy~Nir\altaffilmark{3},
Sagi~Ben-Ami\altaffilmark{1},
Yossi~Shvartzvald\altaffilmark{1},\\
Nora L. Strotjohann\altaffilmark{1},
Enrico~Segre\altaffilmark{2},
Arie~Blumenzweig\altaffilmark{1},
Michael~Engel\altaffilmark{1},
Dennis~Bodewits\altaffilmark{4},
John~W.~Noonan\altaffilmark{4}
}

\altaffiltext{1}{Department of particle physics and astrophysics, Weizmann Institute of Science, 76100 Rehovot, Israel.}
\altaffiltext{2}{Department of Physics Core Facilities, Weizmann Institute of Science, 76100 Rehovot, Israel.}
\altaffiltext{3}{University of California, Berkeley, Department of Astronomy, Berkeley, CA 94720, USA.}
\altaffiltext{4}{Department of Physics, Auburn University, AL 36849, USA.}

\begin{abstract}

Asteroid collisions are one of the main processes responsible
for the evolution of bodies in the main belt.
Using observations of the Dimorphos impact by the DART spacecraft, we estimate how asteroid collisions in the main belt may look in the first hours after the impact.
If the DART event is representative of asteroid collisions with a $\sim1$\,m size impactor, then the light curves of these collisions will rise on time scales of about $\gtrsim100$\,s and will remain bright for about one hour.
Next, the light curve will decay on a few hours time scale to an intermediate luminosity level in which it will remain for several weeks, before slowly returning to its baseline
magnitude.
This estimate suffers from several uncertainties due to, e.g., the diversity of asteroid composition, their material strength, and spread in collision velocities.
We estimate that the rate of collisions in the main belt with energy similar or
larger than the DART impact is of the order of 7000 per year ($\pm1$\,dex).
The large range is due to the uncertainty in the abundance of $\sim1$-m size asteroids.
We estimate the magnitude
distribution of such events in the main belt, and
we show that $\sim6$\% of these events
may peak at magnitudes brighter than 21.
The detection of these events requires a survey
with $\lesssim1$\,hr cadence and may contribute to our understanding of
the asteroids' size distribution, collisional physics,
and dust production.
With an adequate survey strategy, new survey telescopes may regularly detect
asteroid collisions.

\end{abstract}


\section{Introduction}

\label{sec:intro}

Collisions between asteroids are one of the
main processes responsible for shaping the numbers,
size distribution, spins, and dynamical evolution
of objects in the main belt
(\citealt{Dohnanyi+1969JGR_AsteroidsCollisionModel_SizeDistribution, Holsapple+Housen2019P&SS_AsteroidsCollisions_Review, Holsapple2022P&SS_MainBeltAsteroids_CollisionsHistory_Ejecta_Qstar, Masiero+2015aste.book_AsteroidFamilyPhysicalProperties}).
Furthermore, collisions are likely responsible for the replenishment of some of the zodiacal dust (\citealt{Durda+1997Icar_AsteroidCollisions_ZodiacalDustProduction}).

Asteroid collisions have likely already been detected several times:
Prominent examples include the (596) Scheila outburst (\citealt{Jewitt+2011ApJ_596_Scheila_AsteroidCollision_HST_observations, Moreno+2011ApJ_596_Sceila_AsteroidCollision_Observations, Masateru+2011ApJ_596_Scheila_AsteroidCollision_Observations, Bodewits+2011ApJ_Asteroid_696_Scheila_CollisionObservations}), and the P/2010A2 outburst (\citealt{Jewitt+2010Natur_AsteroidCollision_P2010A2}; see however \citealt{Agarwal+2012epsc.conf_P2010A2_ImpactOrRotationalBreakup}).
Many additional cases of asteroid activities were reported, some of which may be due to collisions (e.g., \citealt{Jewitt+2022_TheAsteroidsCometContinuum_AsteroidActivity, Chandler+2023RNAAS_NewActiveAsterod_2015VA108, Chandler+2023RNAAS_NewActiveAsteroid_588045}).

The most studied event is likely the (596) Scheila outburst.
The event was discovered by \cite{Larson+2010CBET_596Scheila_OutburstDiscovery}, likely about one week after the impact (e.g., \citealt{Masateru+2011ApJ_596_Scheila_AsteroidCollision_Observations}).
The slow ejecta mass was estimated to be of the order of $10^{7}$ to a few times $10^{8}$\,kg (e.g., \citealt{Jewitt+2011ApJ_596_Scheila_AsteroidCollision_HST_observations, Hsieh+2012ApJ_OpticalObs_596Scheila_AsteroidCollision, Masateru+2011ApJ_596_Scheila_AsteroidCollision_Observations}) expanding at a velocity of the order of 50\,m\,s$^{-1}$ (e.g., \citealt{Moreno+2011ApJ_596_Sceila_AsteroidCollision_Observations}).
It was estimated this event was produced by a small asteroid, several tens of meter size, colliding with Scheila at a velocity of about 5\,km\,s$^{-1}$.

One shortcoming of the previous detections is that they were likely caught days or weeks
after the impact when the slow-moving ejecta dominates the light from the event.
Based on the known collision events, \cite{Bodewits+2011ApJ_Asteroid_696_Scheila_CollisionObservations} estimated
that collisions, as bright as the Sheila event, occur in the main belt about once every 5 years.
Therefore, it is likely that minor collisions are much more frequent and can be detected regularly.

\cite{McLoughlin+2015Icar_AsteroidCollision_BrightnessIncreasePrediction} used the cratering-size laws of \cite{Housen2007LPI_Asteroid_Collisions_StrengthRockyAsteroids} and \cite{Housen+Holsapple2011Icar_Asteroid_Collisions_EjectaFromImpactCrater} to estimate the brightness increase in the weeks after the impact.
Their study implies a low probability of detection of asteroid collisions using surveys like Pan-STARRS (\citealt{Chambers+2016_PS1_Surveys}) and Catalina Sky Survey (\citealt{Christensen+2019EPSC_CatalinaSkySurvey_Asteroids_NEO_20years}). This is mainly due to the expectation that the week-long transients expected from asteroid collisions are too faint.

Recently, the DART mission (\citealt{Cheng+2018P&SS_DART_mission_Didymos_DeflectionMission, Rivkin+2021PSJ_DART_MissionRequirments})
gave us a glimpse at what an asteroid collision may look like immediately after the impact (\citealt{Daly+2023Nature_DART_KineticImpact,Li+2023Nature_DART_EjectaActiveAsteroid, Graykowski+2023Nature_DART_LightCurve,Shestakova+2023Icar_DART_Impact_Spectroscopy_EmissionLines, Ofek+2024MNRAS_DART_LAST_Swift_Observations}).
Another example was the {\it Deep Impact} collision with comet 9P/Tempel (e.g., \citealt{AHearn+2005Sci_DeepImpact_Tempel1_Comet_ObservationsReview, Ahearn+Combi2007Icar_DeepImpact_Tempel1_Collision}).
The $B_{\rm p}$-band light curve of the DART impact event rises by about 2 magnitudes on a time scale of 100\,s (\citealt{Ofek+2024MNRAS_DART_LAST_Swift_Observations}).
After about ten minutes, the light curve declines to an intermediate level, and after one day,
it starts to decline, on a few days' time scale, to its baseline magnitude.
In \cite{Ofek+2024MNRAS_DART_LAST_Swift_Observations}, we
argued that the finite rise time is due to a transition from an optically thick to an optically thin phase, while the decline represents the time in which some of the ejecta exits the photometric aperture in which the flux is measured.
The DART-impact light curve can be explained by ejecta with two main velocity components.
Fast ejecta moving at 1.6\,km\,s$^{-1}$ (\citealt{Shestakova+2023Icar_DART_Impact_Spectroscopy_EmissionLines, Ofek+2024MNRAS_DART_LAST_Swift_Observations}),
and slow ejecta moving at velocities of up to $\sim10$\,m\,s$^{-1}$ (e.g., \citealt{Li+2023Nature_DART_EjectaActiveAsteroid, Moreno+2023PSJ_DART_Impact_Ejecta_Observations_and_Models, Jewitt+2023ApJ_DART_Impact_DimorphosBoulderSwarm, Ofek+2024MNRAS_DART_LAST_Swift_Observations}).
\cite{Shestakova+2023Icar_DART_Impact_Spectroscopy_EmissionLines} found that the fast-ejecta light is dominated by emission lines
of neutral species (\ion{Na}{1} $\lambda$ 5890, 5896\,\AA,~\ion{K}{1} $\lambda$ 7699\,\AA,~\ion{Li}{1} $\lambda$ 6708\,\AA).
The slow ejecta, on the other hand, are composed
of reflecting $\mu$m to meter size particles (e.g., \citealt{Li+2023Nature_DART_EjectaActiveAsteroid, Jewitt+2023ApJ_DART_Impact_DimorphosBoulderSwarm, Moreno+2023PSJ_DART_Impact_Ejecta_Observations_and_Models, Kim+Jewitt2023arXiv_DART_Impact_SIngleEjectionToExplainDoubleTail, Polishook+2023_DART_Impact_NearIR_spectra, Ofek+2024MNRAS_DART_LAST_Swift_Observations}).
While the fast ejecta were bright in the UV and visible bands, they were not detected in the $z$ band (\citealt{Ofek+2024MNRAS_DART_LAST_Swift_Observations}).
This suggests that the $z$-band does not contain any prominent emission lines.

The DART-Dimorphos impact velocity was about 6.1\,km\,s$^{-1}$ (\citealt{Cheng+2023Nature_DART_MomentumTransfer}), similar to the expected mean asteroid collision velocity of about 6\,km\,s$^{-1}$ (\citealt{Farinella+Davis1992Icar_AsteroidsCollisionsVelocity, Holsapple+Housen2019P&SS_AsteroidsCollisions_Review}).
The DART impactor mass was 579.4\,kg (\citealt{Cheng+2018P&SS_DART_mission_Didymos_DeflectionMission,Cheng+2023Nature_DART_MomentumTransfer}), which is presumably the typical mass of an asteroid with a radius of 0.3-m (assuming a density of about 2.2\,g\,cm$^{-3}$). Impactors with higher mass will likely produce more ejecta and therefore brighter events (e.g., \citealt{Petit+1993_AsteroidsCollisions_Cratering_Qstar_Models, Holsapple+Housen2019P&SS_AsteroidsCollisions_Review, Holsapple2022P&SS_MainBeltAsteroids_CollisionsHistory_Ejecta_Qstar}).

%We therefore assume that the outcome of the DART impact is typical for asteroid collisions. We estimate how such collisions will look when situated in the main belt and make rough predictions for the visibility of asteroid collisions.
%
We use the DART-Dimorphos impact light curve
to show that asteroid collisions in the main belt will produce a $\sim1$\,hr duration
transient, and we estimate the magnitude distribution of these events.
We also estimate the relevant collision and detection rate for such events,
and discuss their detectability and possible survey strategies to search for these collisions.
Detecting asteroid collisions
shortly after the impact,
will allow us to compare these events with the DART impact, 
study asteroid collision physics, surface weathering,
estimate the frequency of meter-sized asteroids, and evaluate the zodiacal dust production rate from asteroid collisions. 
The calculations in this paper were performed using tools available\footnote{\url{https://github.com/EranOfek/AstroPack} tag v1.1} in \cite{Ofek2014_MAAT,Ofek+2023PASP_LAST_PipeplineI}.

In \S\ref{sec:ScalingLC} we scale the light curve of a DART-like impact to the distance of the main belt,
while the expected magnitude distribution of asteroid collisions is estimated in \S\ref{sec:MagDist}.
In \S\ref{sec:Rate} we estimate the collision rate in the main belt,
while survey strategies are discussed in \S\ref{sec:Survey},
and we conclude in \S\ref{sec:Disc}.

\section{Scaling the DART light curve to the Main Belt}
\label{sec:ScalingLC}

The uncertainties regarding the outcome of asteroid collisions are large and depend on multiple parameters like 
the material strength, structure, impact angle, and more (e.g., \citealt{Dohnanyi+1969JGR_AsteroidsCollisionModel_SizeDistribution, Gault+1969JGR_AsteroidsCollisionsDistruction_Qstar, Luther+2018M&PS_AsteroidsCollisions_DART_ImpactModeling, Holsapple+Housen2019P&SS_AsteroidsCollisions_Review, Holsapple2022P&SS_MainBeltAsteroids_CollisionsHistory_Ejecta_Qstar, Statler+2022arXiv_DART_Didymos_PredictionsSimulations}).
Another important parameter are the masses of the impactor and target.
The asteroid size distribution, even at small sizes, likely follows a steep power law (e.g., \citealt{Dohnanyi+1969JGR_AsteroidsCollisionModel_SizeDistribution,Pan+Sari2005_KBO_SizeDistribution_MaterialStrength}).
Therefore, the collision rate is dominated by the smallest impactors and targets.
Given this fact, we will hereafter use the approximation that all collisions
between asteroids larger than $\sim1$\,m,
generate as bright events as the DART event.
In any case, our results
only provide an order-of-magnitude estimate of asteroid collision visibility.
In addition, our rate estimate (\S\ref{sec:Rate}) suffers from major uncertainties
due to the unknown size distribution, and numbers, of small ($\sim1$\,m) asteroids.

\begin{figure}
\centerline{\includegraphics[width=7.5cm]{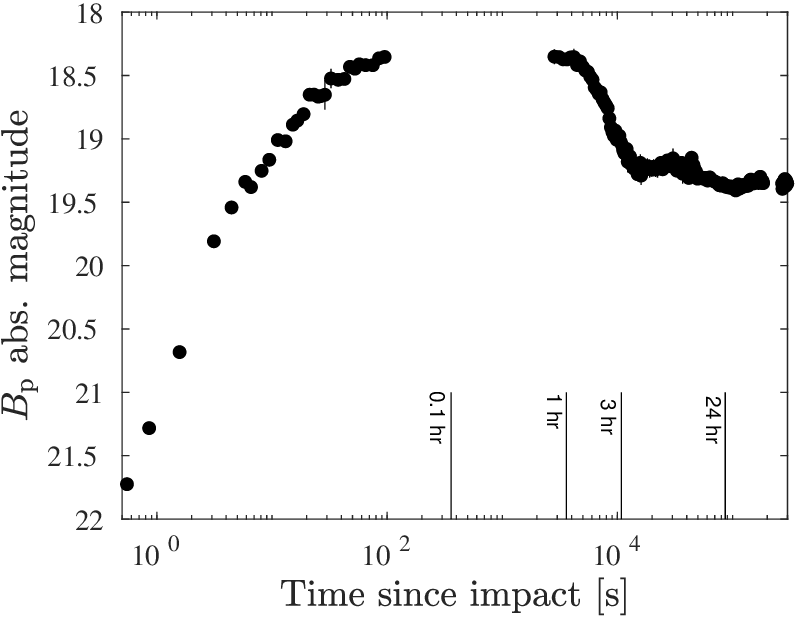}}
\caption{The time-stretched light curve of the Dimorphos-DART impact
shifted to a distance of 1\,AU from the observer,
1\,AU from the Sun and a phase angle of 53.3\,deg. The time stretch is applied to account for
the time it takes the flux to get out of the photometric aperture.
See text for details.
\label{fig:DART_StretchedLC_at2au}}
\end{figure}

We adopt the light curve of the DART-Dimorphos impact as was reported by \cite{Ofek+2024MNRAS_DART_LAST_Swift_Observations}.
This data set was obtained using the Large Array Survey Telescope (LAST; \citealt{Ofek+BenAmi2020_Grasp_SkySurvrys_CostEffectivness, BenAmi+2023PASP_LAST_Science})
and produced by the LAST data reduction pipeline (\citealt{Ofek+2023PASP_LAST_PipeplineI, Ofek2014_MAAT, Soumagnac+Ofek2018_catsHTM, Ofek2019_Astrometry_Code}).
LAST observations were taken without a filter, in a band covering the $\sim4000$\,\AA~to $\sim8000$\,\AA~and calibrated to the GAIA $B_{\rm p}$ band (\citealt{GAIA+2016_GAIA_mission, Evans+2018_GAIADR2_photometry, GAIA+2021_GAIAEDR3_Summary_Content}).
Here, we adopt the LAST light curve obtained using an aperture radius of $12.5''$.
Figure~\ref{fig:DART_StretchedLC_at2au} shows the Dimorphos-DART light curve if it had occurred at a distance of 1\,AU
from the observer and 1\,AU from the Sun, with a phase angle of $\theta=53.3$\,deg.
%This observing geometry is consistent with a small asteroid impact on a Aten ($a<$1 au, $Q>$0.983 au)  or Apollo ($a>$1 au, $q<$1.017 au) asteroid.
%
We assume that the DART impact peaked at a Vega\footnote{The conversion between AB magnitude to Vega magnitude in the $B_{\rm p}$ band is $m_{\rm AB}=m_{\rm Vega}+0.0155$.} magnitude of 12.8
in the GAIA-$B_{\rm p}$ band (after removing the Didymos pre-impact light).
The DART impact took place at a phase angle of $\theta=53.3$\,deg,
a solar distance of $r_{\rm DART}\cong 1.05$\,AU, and Geocentric distance of $\Delta_{\rm DART}\cong 0.074$\,AU.
We also assume that the
baseline contribution of the target and impactor asteroids are negligible
(i.e., the baseline flux of Didymos was removed).
No time stretch was applied, for data points up to the peak of the event (at $\approx 100$\,s), while for all times after 100\,s, we stretch the time by a factor of 13 ($\cong1\,AU/\Delta_{\rm DART}$; i.e., the ratio of the observer distance to the DART-impact distance).
The reason is that in early times the light curve rise is dominated by the transition from optically thick to optically thin scattering, which is independent of the distance to the observer.
However, when the material becomes optically thin (during maximum light), the surface area of the scattering particles is constant
and the flux will start to decline only when the expanding ejecta become resolved, and some of the flux exits the photometric aperture.
A possible caveat is that emission lines produced by neutral species with short lifetimes are present in the aperture.
For example, \cite{Shestakova+2023Icar_DART_Impact_Spectroscopy_EmissionLines} estimated that at a distance of 1\,AU from the Sun\footnote{This time scale increases with the distance from the Sun.}, \ion{Na}{1} and \ion{K}{1} will become ionized on a time scale of 1\,day, while \ion{Li}{1} will become ionized within about 1\,hr.
Since, at least in the case of the DART impact, \ion{Na}{1} and \ion{K}{1} line dominate the emission, we do not expect significant variations in the total emitted light\footnote{By total emitted light we mean the light measured in infinite photometric aperture.}, in the first several hours after the impact.

In \cite{Ofek+2024MNRAS_DART_LAST_Swift_Observations}, we present an optical-thickness model for the rise time of the DART-impact
light curve. This rise-time is roughly given by
\begin{equation}
    t_{0} \sim \sqrt{\frac{9 m_{\rm ej} \sigma_{\rm T} }{8\pi^2 \rho r_{\rm p}^3  v_{\rm ej}^{2}} }.
    \label{eq:t0_mono}
\end{equation}
Here, $m_{\rm ej}$ is the ejecta mass,
$\sigma_{\rm T}$ is the total absorption and scattering cross-section of the particles in the ejecta, $r_{\rm p}$ is the cross-section radius of the particles, and
$v_{\rm ej}$ is the ejecta velocity.
The rise time, therefore, may provide some information on the ejecta mass, which presumably depends on the impactor mass.
However, $m_{\rm ej}$, $v_{\rm ej}$, $r_{\rm p}$ and $\rho$ are uncertain,
and therefore, a robust estimation of the mass is currently not possible.

The relationship between the properties of the light curve and the phase angle remains uncertain. Specifically, if we assume that the emission from the fast ejecta is primarily governed by emission lines resulting from resonant scattering (\citealt{Shestakova+2023Icar_DART_Impact_Spectroscopy_EmissionLines}), the geometry of the ejecta could influence the absorption probability. Consequently, this variation may impact the overall brightness of the event. However, we do not anticipate a substantial dependence on the phase angle unless some of the light is due to reflectance from small particles.
For the small particles ($\sim1\,\mu$m) in the slow ejecta, phase angle-dependent Mie scattering dominates.
For larger particles, collective effects like
coherent backscattering and shadow may become important (e.g., \citealt{Hapke1986Icar_OppositionEffect_Extinction_Theory, Hapke+1993Sci_OppositionEffect_BackScattering}).
Since the prompt visibility of 1-m size asteroid collisions is presumably dominated by the fast ejecta, we will ignore phase angle effects.
Due to a lack of additional information,
we assume that the DART impact scattering
properties are representative of asteroid collisions.

\section{The apparent magnitude distribution of collision events}
\label{sec:MagDist}

To calculate the magnitude distribution of collision events, we need the heliocentric ($r$) and Geocentric ($\Delta$) distance distribution of asteroids in the main belt.
Generating ephemerides to known numbered asteroids at random times results in a non-uniform ecliptic longitude distribution with peak-to-valley difference, relative to the mean, of about 70\%.
To obtain the mean magnitude distribution, we create a mock catalog of asteroids by generating ephemerides for all known
numbered asteroids\footnote{Orbital elements from \url{http://ssd.jpl.nasa.gov/dat/ELEMENTS.NUMBR.gz}} ($\approx6\times10^{5}$) between 2000 Jan 1 till 2001 Jan 1, in steps of 10 days.
This resulted in a mock catalog with about $2.2\times10^{7}$ asteroids (i.e., the position of each asteroid is evaluated at multiple times).
This reduces the variations in the ecliptic longitude distribution to about\footnote{As we increase the catalog size by a factor of about 36 we expected a decrease in the variations by about $\sqrt{36}$.} 10\%.
We included in the catalog only objects with semi-major axis in the range of 1.7 to 5.5\,AU.
About 63\% of the asteroids in this mock catalog have Sun-observer-asteroid angle larger than 50\,deg.

\begin{figure}
\centerline{\includegraphics[width=7.5cm]{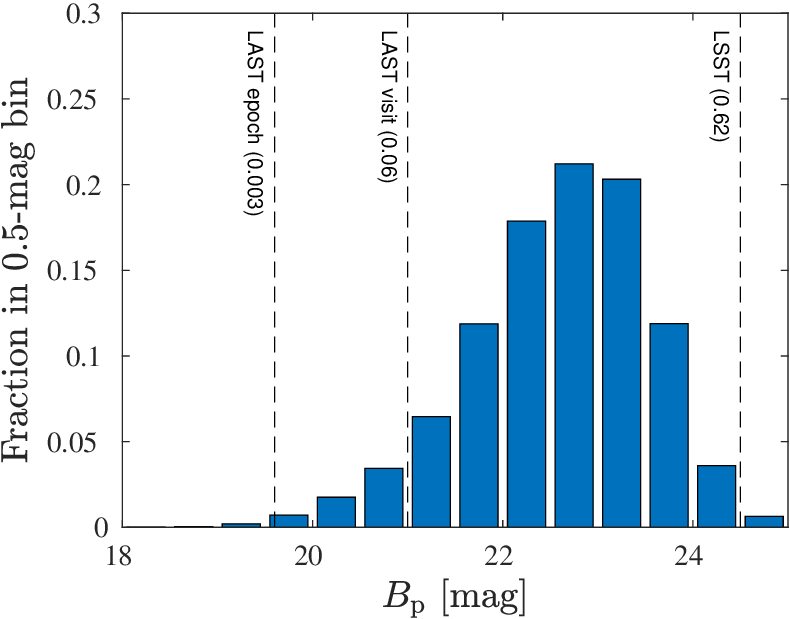}}
\caption{The normalized magnitude distribution of the expected DART-like impacts in the main belt,
with Sun-Observer-Asteroid angle larger than 50\,deg.
The dashed lines show the limiting magnitude
of the LAST
epoch (20\,s exposure), LAST visit ($20\times20$\,s exposure), and the Large Survey of Space and Time (LSST; \citealt{Ivezic+2019_LSST_Survey}) single image.
The fraction of events brighter than the limiting magnitude is given in parentheses.
These fractions are already multiplied by the fraction of the asteroid with
a Sun-Observer-Asteroid angle larger than 50\,deg ($0.63$).
\label{fig:Sim_MagDist}}
\end{figure}

\begin{figure}
\centerline{\includegraphics[width=7.5cm]{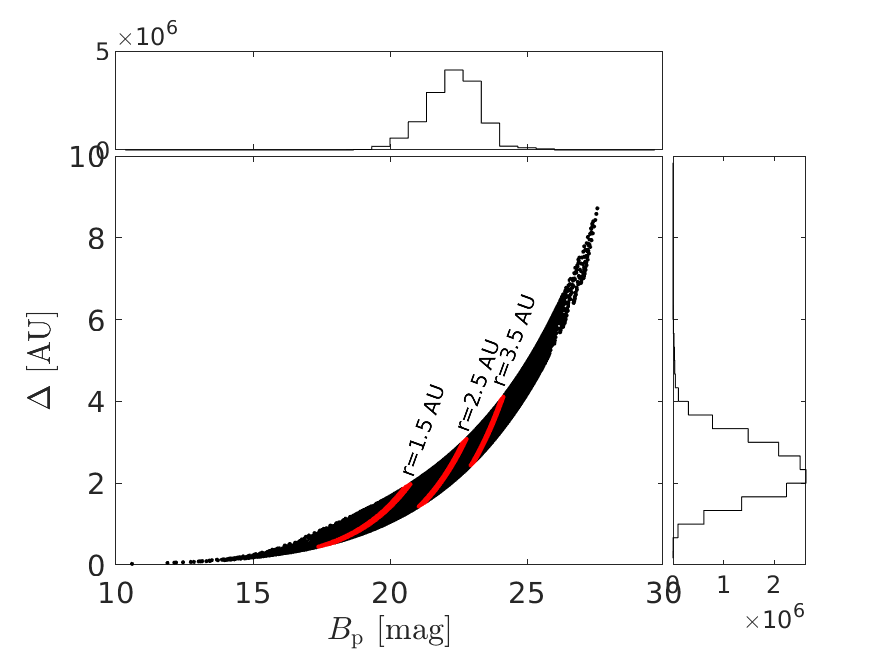}}
\caption{The magnitude vs.~$\Delta$ of expected DART-like impacts in the main belt,
which have Sun-Observer-Asteroid angles larger than 50\,deg.
Asteroids with constant Heliocentric distance ($r$) of 1.5, 2.5, and 3.5\,AU
are marked in red.
\label{fig:Sim_Mag_Delta}}
\end{figure}

\begin{figure}
\centerline{\includegraphics[width=7.5cm]{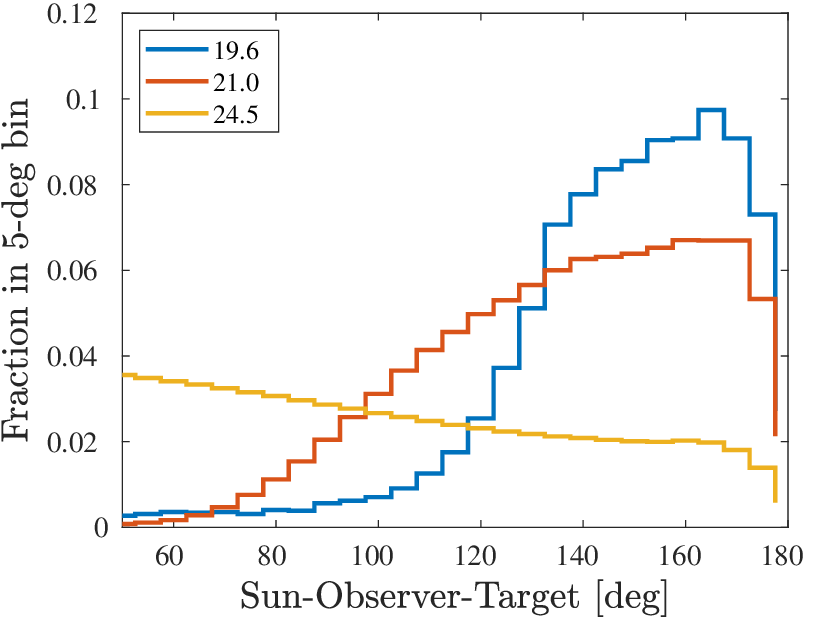}}
\caption{The probability distribution of the Sun-observer-target angle (elongation) for events brighter than magnitude 19.6 (blue line), 21.0 (magenta line), and 24.5 (yellow line).
The drop near 180\,deg is due to the non-zero ecliptic latitude of asteroids.
\label{fig:Sim_SOT}}
\end{figure}

For each simulated asteroid,
we scale the peak $B_{\rm p}$ magnitude of the DART impact (12.6\,mag), according to its heliocentric ($r$) and geocentric ($\Delta$) distances.
The expected $B_{\rm p}$ magnitude distribution,
of events with 
Sun-Observer-Asteroid angle larger than 50\,deg,
is shown in Figure~\ref{fig:Sim_MagDist}.
The dashed vertical lines represent limiting magnitudes of 19.6, 21.0, and 24.5.
In parenthesis, next to each line,
we give the estimated fraction of events brighter than that marked by the corresponding line (i.e., the cumulative sum of the histogram was multiplied by $0.63$, which is the fraction of asteroids with Sun-Observer-Asteroid angle larger than 50\,deg.).
In Figure~\ref{fig:Sim_Mag_Delta}, we show the magnitude vs.~$\Delta$
distribution.
Figure~\ref{fig:Sim_SOT} shows the probability distribution of the
Sun-Observer-Asteroid angle for collision events brighter than a magnitude of 19.6 (blue), 21.0 (magenta), and 24.5 (yellow).

The time, from impact, it takes to the light curve of the event to start decaying is roughly equal to the time it takes the ejecta to cross the photometric aperture. 
For a DART-like impact, with ejecta velocity of the order of 10\,m\,s$^{-1}$,
and photometric aperture of about $10''$, this time scale is of the order of:
\begin{equation}
    t_{\rm dur}\sim 2600 \frac{\Delta}{1\,{\rm AU}}\,{\rm s}.
\end{equation}
This relation was normalized such that the duration
of the DART impact at $\Delta=0.074$\,AU is about 200\,s matching the duration observed by \cite{Ofek+2024MNRAS_DART_LAST_Swift_Observations}.
Collision simulations suggest that
the fastest parts of the ejecta
from asteroid collisions move
at velocities of the order of 0.1 to 0.4
of the impact velocity (e.g., \citealt{Luther+2018M&PS_AsteroidsCollisions_DART_ImpactModeling}).
If accurate, this range would introduce about a half-order of magnitude uncertainty in the duration
of the events.

We note that in the Scheila (596) event, the magnitude of the extended dust cloud was about
14.4\,mag which corresponds to an absolute planetary magnitude\footnote{Defined as the magnitude at $\Delta=1$\,AU, $r=1$\,AU, and at opposition, although in the Scheila case, the Sun-Observer-Asteroid angle was about 115\,deg when it was discovered.} of about 10.
This is more than 8 magnitudes brighter than the fast-ejecta brightness of the Dimorphos-DART impact. This suggests that the Scheila (596) event was generated by an impactor with a mass much larger than that of the DART spacecraft.
However, as estimated by \cite{Bodewits+2011ApJ_Asteroid_696_Scheila_CollisionObservations}
such events are relatively rare -- about one event every five years.

The DART impact was also detected using {\it Swift}-UVOT (\citealt{Gehrels+2004_Swift, Roming+2005SSRv_Swift_UVOT}) in the
near ultraviolet (NUV) band (\citealt{Ofek+2024MNRAS_DART_LAST_Swift_Observations}),
peaking at about 16.4\,mag.
Such events in the $NUV$ band can be potentially detected by {\it ULTRASAT} (\citealt{Sagiv+2014_ULTRASAT, Shvartzvald+2023PASP_ULTRASAT_Overview}).
Therefore, we also calculate the expected magnitude distribution in the $NUV$ band and find that the fraction of collision events at Sun-Observer-Asteroid angles larger than 50\,deg, and brighter than the limiting magnitude of {\it ULTRASAT} in 900\,s, is about $10^{-6}$.
This suggests that the rate of such events in the $NUV$ is low.

We also estimate the fraction of events for a TESS-like observatory (\citealt{Ricker+2015_TESS_MissionInstrument}).
For a limiting magnitude of 17.8 in 200\,s, the fraction
of events that may be detected is about $8\times10^{-5}$.

\section{Collision rate}
\label{sec:Rate}

The collision rate of asteroids in the main belt is given by:
\begin{equation}
    R_{\rm coll} \approx P_{i} R_{\rm t}^{2} N_{\rm t}(>R_{\rm t}) N_{\rm imp}(>R_{\rm imp}).
    \label{eq:RateN}
\end{equation}
Here $R_{\rm t}$ is the target radius, $R_{\rm imp}$ is the impactor radius,
$N_{\rm t}(>R_{\rm t})$ is the number of asteroid targets with radius larger than $R_{\rm t}$,
$N_{\rm imp}(>R_{\rm imp})$ is the same for the impactor,
and $P_{i}$ is the mean {\it intrinsic collision probability} defined such that the number of collisions between two particles occurrs in a time interval $\Delta{t}$ is $P_{i}(R_{\rm t}+R_{i})^{2}\Delta{t}$.
\cite{Farinella+Davis1992Icar_AsteroidsCollisionsVelocity} used simulations to estimate the average intrinsic collision probability in the main belt:
$P_{i}=2.85\times10^{-18}$\,km$^{-2}$\,yr$^{-1}$.
This value was estimated based on a sample of 682 asteroids with diameter $>50$\,km. Therefore, we assume that this collision probability is also valid for smaller asteroids.
The value of $P_{i}$ depends, in order of importance, on the semi-major axis, inclination, and eccentricity.
However, given that the main uncertainty in our estimate is due to the poorly constrained number of small asteroids, here we use the average intrinsic collision probability.
We note that we tested the effect of the semi-major axis dependency of $P_{i}$
on the magnitude distribution in Figure~\ref{fig:Sim_MagDist} and found that it typically changes this plot by less than 10\%.
Additional caveats in the estimation of $P_{i}$ and the average collision speeds between asteroids are discussed in \cite{Farinella+Davis1992Icar_AsteroidsCollisionsVelocity}.

Assuming target and impactor densities of about 2.2\,g\,cm$^{-3}$,
an impactor with a kinetic energy equivalent to DART would have a radius $R_{\rm imp}\approx0.3$\,m.
From energy conservation, we constrain the the DART impact's
fast ejecta mass to $\lesssim10^{4}$\,kg,
and therefore we need $R_{\rm t}\gtrsim1.1$\,m.
These requirements are based on very simplistic assumptions
and in reality they can be different, e.g., due to the uncertainties in collision physics.

The cumulative number of
asteroids is usually
parameterized as a power law $N(>r)\propto r^{-q}$.
However, the power-law index
is size-dependent and also asteroid type-dependent (e.g., \citealt{Yoshida+2007P&SS_AsteroidSurvey_Subaru_HSC, Maeda+2021AJ_AsteroidSizeDistribution_Subaru_HSC, Garcia-Martin+2023DPS_FaintAsteroidsMagDist_HST}).
The cumulative number of asteroids larger than about 0.4\,m is not measured directly.
\cite{Maeda+2021AJ_AsteroidSizeDistribution_Subaru_HSC}
estimated that about $10^{7}$ main-belt asteroids have radii larger than 150\,m.
They also found that the average power law index, $q$, is about $2.75$
for asteroids in the radius range of 150\,m to 500\,m. 
Rewriting Equation~\ref{eq:RateN}
with the power-law size distribution gives:
\begin{equation}
    R_{\rm coll} \approx P_{i} R_{\rm t}^{2} C_{\rm r}^{2} R_{\rm t}^{-q} R_{\rm imp}^{-q},
\end{equation}
where, for radius measured in km and assuming an albedo of $0.12$, $C_{\rm r}\approx5.4\times10^{4}$ (Figure~10 in \citealt{Maeda+2021AJ_AsteroidSizeDistribution_Subaru_HSC}).
Given these values, 
and assuming that the power-law index, $q=2.75$, 
is valid down to the asteroid size of $\sim0.3$\,m;
$R_{\rm t}\approx1.1$\,m; and $R_{\rm imp}\approx0.3$\,m,
we find a main-belt asteroids collision rate of $\sim7000$\,yr$^{-1}$.
If one assumes that the power-law can be extrapolated, the uncertainty on
the power-law index, of about 5\%, is translated to a factor of about 2 uncertainty on the number of $\sim1$\,m size asteroids.
% 1e7.*(0.0004./0.15).^-2.75 .*1e7.*(0.001./0.15).^-2.75.*Pi.*0.001.^2
Another major uncertainty is related to $R_{\rm t}$ and $R_{\rm i}$ that are needed in order to produce a DART-like display.
Therefore, the uncertainty of the collision rate is likely about an order of magnitude in each direction.

\section{Survey strategies and collision detectability}
\label{sec:Survey}

Here, we discuss the requirements for a survey that can detect asteroid collisions efficiently.
Given the expected short duration of these events ($\sim0.5$--$3$\,hr), a relatively fast cadence of $\lesssim1$\,hr is crucial.

%An important question for survey design is: given the expected magnitude distribution of the collision events, what is the limiting magnitude that will result in the fastest survey speed (i.e., highest asteroid collision detection rate)?
%
%Assuming the survey is operating in the background-dominated noise regime (e.g., \citealt{Ofek+BenAmi2020_Grasp_SkySurvrys_CostEffectivness}), the limiting flux of the system goes like $t_{\rm exp}^{-1/2}$, where $t_{\rm exp}$ is the exposure time.
%Approximating, the cumulative rate of events per unit angular area ($\dot{n}[>F]$) as a function of flux $F$, as a power law, with a power-law index $\alpha$, we can write $\dot{n}(>F)\sim F^{-\alpha}$. In this case, the observed rate of events is $\dot{n}\propto t_{\rm exp}^{\alpha/2 -1}$. Therefore, $\dot{n}$ is maximal when $t_{\rm exp}^{\alpha/2 -1}$ is maximal.
%
%Therefore, we would like the limiting magnitude of the search to correspond to the flux at which the cumulative magnitude distribution has the largest positive derivative.
%
%From Figure~\ref{fig:Sim_MagDist}, we see this happens around magnitude of $\sim19.5$. However, a deeper limiting magnitude is important in order to avoid marginal detections and to better characterize the detected events.

An efficient survey needs to cover the ecliptic region.
Since, at any given moment, about 95\% of the known main-belt asteroids are found within 20\,deg of the ecliptic,
covering $\pm20$\,deg of the ecliptic is sufficient.
Furthermore, assuming no strong phase-angle dependencies, Figure~\ref{fig:Sim_SOT} suggests that shallow surveys can detect only the brightest (nearest) events. Therefore, observing at elongations (i.e., angular separation between Sun and target)
$\gtrsim 120$\,deg is most efficient. However, for deep surveys like LSST, most of the events will be detected at low elongations.
%\textbf{DP - another complexity: outer belt has asteroids with lower albedo, so there ejecta might be darker, and it will be harder to detct them, compared to inner main belt asteroids that have higher albedos.} - NOT relevant, as the emission is via lines.

Given the expected high rate of asteroid collisions,
of about $1$ events per day brighter than magnitude 21,
an important question is why such events, with $\sim1$\,hr time scale, were not found in the past.
Typically, sky surveys 
(e.g., \citealt{Heinze+2018_ATLAS_VarStars, Chambers+2016_PS1_Surveys, Bellm+2019_ZTF_Overview})
require
at least two epochs in a single band, to identify astrophysical transients.
This is required in order to
reject imaging artifacts (e.g., cosmic rays),
and satellite glints (e.g., \citealt{Corbett+2020_SatellitesGlints, Nir+2020_Satellites_Glints_FlaresLimit, Nir+2021_RNASS_GN-z11-Flash_SatelliteGlint}).
Furthermore, these images needed to be separated in time
to screen for unknown
Solar System objects.
However, asteroid collisions with light curves similar to those shown in Figure~\ref{fig:DART_StretchedLC_at2au}, will in most cases only be detected in less than four images, and they will not be recovered in the next days
after the first detection.
Therefore, using current typical cadences, most asteroid collision events will be missed.
In that respect, the LAST
survey strategy is an important advantage.
Specifically, LAST observes every field with $20\times20$\,s exposure, called a visit. This visit strategy allows for
cosmic ray and satellite glint mitigation, 
as well as for detecting the motion of most main belt asteroids in a single visit.
Furthermore, LAST has the potential, in some cases, to detect flux variability in a single visit (6\,min time scale).
The probability of LAST capturing an event will depend on the exact survey strategy. 
For example, if the LAST high cadence survey of about 2000\,deg$^{2}$ will be conducted mainly near the ecliptic, then LAST may be able to detect collision events at a rate of the order of ten per year.

Another channel to discover asteroid collisions is via their extended emission and the tails they develop.
Based on the DART experience, the brightness of the extended emission and tail is lower than the magnitude of the prompt brightening, which is the subject of this paper. For example, the brightness of the extended phase of the DART impact is a factor of two ($0.75$\,mag) below that of the prompt phase. 
Furthermore, the mass of the slow-ejecta in the DART impact is in the range of $10^6$ to $10^7$\,kg (e.g., \citealt{Kim+Jewitt2023arXiv_DART_Impact_SIngleEjectionToExplainDoubleTail, Ofek+2024MNRAS_DART_LAST_Swift_Observations}) – this is between two to three orders of magnitude higher, than the conservation-of-energy upper limit on the mass of the fast ejecta.
Releasing this amount of ejecta requires an asteroid whose radius is at least a factor of 5 larger than our assumed $R_{\rm t}$ (Eq.~\ref{eq:RateN}). Assuming $q\approx2.75$, this means that the rate of events that produces slow ejecta mass similar to that produced in the DART event, is at least two orders of magnitude smaller (i.e., $\sim 5^{-2.75}$),  compared to the rate we estimate for the prompt events.
Assuming that the currently deepest surveys that cover a large fraction of the sky (e.g., Pan-STARRS; \citealt{Chambers+2016_PS1_Surveys}) is complete to a limiting magnitude of about 21\footnote{See transients detection magnitude distribution in \url{https://www.wis-tns.org/stats-maps/plots}.},
we expect that a Pan-STARRS-like survey will find $\lesssim1$ asteroid collision events every year (see however, \citealt{McLoughlin+2015Icar_AsteroidCollision_BrightnessIncreasePrediction}).
Since, in the slow ejecta phase, these transients should be visible for at least several weeks, they should be found by ongoing surveys. Interestingly, from 2019 to 2021, on average, about 3 active asteroids were reported every year (e.g., \citealt{Jewitt+2022_TheAsteroidsCometContinuum_AsteroidActivity}). Although there are other reasons for asteroids to become active (e.g., rotational breakup), at least some of these detections may be due to collisions of small (10-m size) asteroids. Due to the large uncertainty, in our rate estimate, we conclude that the estimated high rate of collisions, presented in this paper, can~not be ruled out based on existing data.

\section{Discussion}
\label{sec:Disc}

The detection of asteroid collisions in real time, soon after the event, can be used to constrain the size distribution of asteroids and their evolution and provide feedback to collision simulations.
Here, we use the DART impact as a benchmark for such collisions and estimate their observed properties and rate.
We estimate that about 7000 collisions per year,
with a brightness increase similar or larger than the DART impact, occur in the main belt.
Furthermore, we estimate that about 6\% of these events
will be brighter, at peak, than magnitude 21.
Therefore, we expect of the order of $1$ events per day brighter than magnitude 21 and at angular distances $>50$\,deg from the Sun.

Both the rate and visibility estimates are highly uncertain.
The main reasons are the large uncertainty on the number of small ($\sim1$\,m sized) asteroids,
and the possible diversity in asteroids' physical properties
that can presumably modify the optical output, compared to the DART impact event.
However, detecting asteroid collisions can reduce some of these uncertainties.
The main goal of our estimate is to guide possible searches and to help design survey strategies.

Although bright asteroid collisions are presumably very common,
we speculate that the main reasons that such events were
not found previously by sky surveys such as PTF, PanSTARRS, ATLAS, and ZTF (e.g.,
\citealt{Law+2009_PTF, Chambers+2016_PS1_Surveys, Heinze+2018_ATLAS_VarStars, Bellm+2019_ZTF_Overview}), hours after the impact,
include:
(i) the requirement of transient surveys for at least two detections separated by $\sim1$\,hr at the same sky location;
and (ii) the requirement of asteroid surveys for (typically) four or more measurements over a time scale of days.
These problems can be avoided using dedicated surveys, or awareness by the survey analysis team to the possibility
of finding such events.

\section*{Acknowledgements}

We thank an anonymous referee for useful comments on the manuscript.
E.O.O. is grateful for the support of
grants from the 
Benozio center,
Willner Family Leadership Institute,
Ilan Gluzman (Secaucus NJ), Madame Olga Klein - Astrachan,
Minerva foundation,
Israel Science Foundation,
BSF-NSF, Israel Ministry of Science,
Yeda-Sela, Sagol Weizmann-MIT, and the
Rosa and Emilio Segr\'e Research Award. D.P. is thankful to the Israeli space agency and their near-Earth asteroids mitigation efforts.
This research is supported by the Israeli Council for Higher Education (CHE) via the Weizmann Data Science Research Center, and by a research grant from the Estate of Harry Schutzman.

%\section*{Data Availability}

%The data presented in this paper is available in the electronic tables, while the code is accessible via GitHub\footnote{https://github.com/EranOfek/AstroPack}.

\bibliography{papers.bib}
\bibliographystyle{aasjournal}

\end{document}